\documentclass[twocolumn,twoside,slac_two]{revtex4}
\usepackage{graphicx}
\usepackage{fancyhdr}
\def\ipb{\rm pb^{-1}}
\def\fd{f_{D^+}}
\def\fds{f_{D_s^+}}
\def\ep{\epsilon^{\prime}}
\def\edp{\epsilon^{\prime\prime}}
\def\bf{{\cal{B}}}
\def\dsmunu{D_s^+\to\mu^+\nu}
\def\dstaunu{D_s^+\to\tau^+\nu,~\tau^+\to\pi^+\nu\bar{\nu}}
\def\dstaunue{D_s^+\to\tau^+\nu,~\tau^+\to e^+\nu\bar{\nu}}

\def\etap{\eta^{\prime}}
\def\ecctrk{E_{\rm CC}^{\rm trk}}
\def\mt{m_{\tau^+}}
\def\mm{m_{\mu^+}}
\def\mds{m_{D_s^+}}

\begin{document}

\title{Leptonic $D$ and $D_s$ Decays near $c\bar{c}$ Threshold}

%

\author{S. R. Blusk}
\affiliation{Syracuse University, Syracuse, NY 13244, USA}

\begin{abstract}
We present recent results from the CLEO Collaboration on leptonic decay rates 
of $D$ and $D_s$ near $c\bar{c}$ production threshold. From these decay rates,
we extract the decay constants, $\fd=(222.6\pm16.7^{+2.8}_{-3.4})~{\rm MeV}$,
$\fds = (274\pm10\pm5)~{\rm MeV}$, and the ratio $\fds/\fd=1.23\pm0.11\pm0.03$.
\end{abstract}

\maketitle

\thispagestyle{fancy}


\section{Introduction}

Within the Standard Model, leptonic $D$ (or $B$) meson decays proceed via annihilation of the 
initial state quarks. The matrix element is described by the product of 
a hadronic current, a leptonic current, along with a $W$ propagator. The form of the 
latter two are well-known within the Standard Model, however, the hadronic matrix element,
which represents the annihilation of the initial state heavy quark and light antiquark,
depends on the details of the initial-state quark wave-functions, and is
not calculable using standard techniques of perturbative QCD. This hadronic
matrix element can be computed using either lattice QCD~\cite{lqcd1,lqcd2,lqcd3}, 
or other techniques~\cite{qcdspectral,qcdsr,rqm,potmod,isomassspl}. The partial width
for the leptonic decay is given by:

\begin{equation}
\Gamma(D^+_{(s)}\to l^+\nu)={G_F^2\over 8\pi}f_{D^+_{(s)}}^2 m_l^2 M_{D^+_{(s)}} (1-{m_l^2\over M_{D^+_{(s)}}^2})^2|V_{cd(s)}|^2,
\label{eq:width}
\end{equation}

\noindent where $G_F$ is the Fermi constant, $M_{D^+_{(s)}}$ is the $D^+$ ($D_s^+$) mass, $m_l$ is the final state
lepton's mass, and $V_{cd(s)}$ are the relevant CKM matrix elements, The quantity $f_D$ is the decay constant
and represents the hadronic matrix element discussed above. A critical input to $B$ mixing and CP violation
measurements in the $B$ sector is the $B$ decay constant, $f_B$. Due to the difficulty in measuring
$f_B$, we take the value from theory, usually lattice QCD. To have confidence in the theoretical number,
a stringent theoretical test is provided by a precision measurement of the $D$ decay constant, $f_D$.
Such a measurement provides a critical test of any theory or model that makes predictions for decay constants.

      The CLEO experiment, operating near $c\bar{c}$ threshold, is well positioned to measure these
decay rates, and hence $f_{D^+}$ and $f_{D^+_{s}}$. Charge conjugate finals states are implied
throughout unless otherwise noted.

\section{Measurement of $f_D^+$}
\begin{figure}[h]
\centering
\includegraphics[width=80mm]{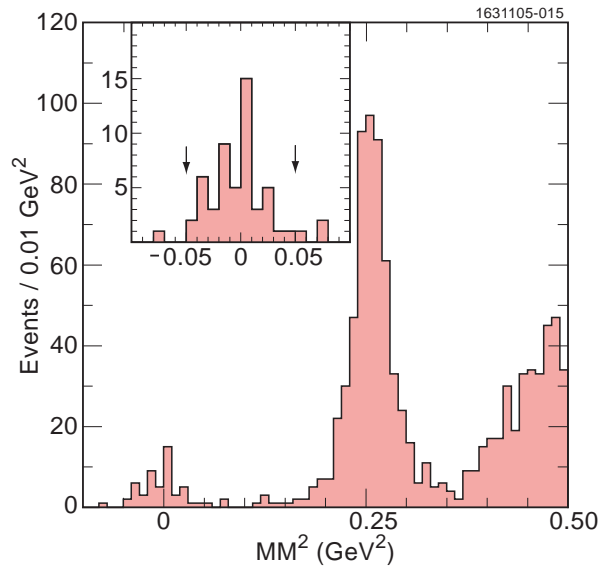}
\caption{Missing-mass squared distribution for $D^+\to\mu^+\nu$ candidates. The peak near zero
corresponds to signal events, and is expanded in the inset. The larger peak at 
$MM^2\simeq0.25$~GeV$^2$ corresponds to $D^+\to K_{S,L}\pi^+$ events
which pass the selection requirements.} \label{fig:dp2munu}
\end{figure}

   To measure $f_D^+$~\cite{dplep}, we use 281~$\ipb$ of data collected at the $\psi(3770)$ resonance.
The proximity to the production threshold implies that the $\psi(3770)$ decays to 
$D\bar{D}$ with no additional particles. We exploit this clean final state, along with the
hermiticity of the detector to {\it reconstruct} the neutrino from the missing momentum in the event.
Specifically, we fully reconstruct a $D^-$ meson (the {\it tag}) in six hadronic final states,
comprising $N_{\rm tag}=158,354\pm496$ tags. To search for $D^+\to\mu^+\nu$, we require a single
extra charged particle with an energy deposition in the crystal calorimeter (CC), 
$\ecctrk<$300 MeV,
and veto events with any additional photon candidates with energy larger than 250 MeV. From this
subsample of events, we compute the square of the missing-mass ($MM^2$) recoiling against the $D^-\mu^+$ system.
For $D^+\to\mu^+\nu_{\mu}$, a peak at zero is obtained with a resolution of $\sigma(MM^2)\sim0.025$~GeV$^2$.
The $MM^2$ distribution is shown in Fig.~\ref{fig:dp2munu} for data. The
clear excess near zero is the $D^+\to\mu^+\nu$ signal. Some $D^+\to K_{S,L}\pi^+$ events
pass the selection requirements and appear as a prominent, but well-separated peak near
$MM^2\simeq0.25$~GeV$^2$.

The branching fraction is computed using:

\begin{equation}
{\cal{B}} = {N_{\rm cand} - N_{\rm back} \over N_{\rm tag}\epsilon_{\mu}\epsilon_{CC} },
\end{equation}

\noindent where $N_{\rm cand}=50$ is the number of signal candidates in the region $|MM^2|<0.050$~GeV$^2$, 
$N_{\rm back} = 2.81\pm0.30\pm0.27$ 
is the expected number of background events, $N_{\rm tag}=158,354\pm496$ is the number of fully-reconstructed
$D^-$ tags, $\epsilon_{\mu}=69.4\%$ is the efficiency for reconstructing and identifying the muon,
and $\epsilon_{CC}=96.1\%$ is the fraction of events that do not have any additional photon candidates
with energy larger than 250~MeV. An additional correction of ($1.5\pm0.4\pm0.5$)\%
is applied to account for the higher efficiency for reconstructing a $D^-$ tag in $D^+\to\mu^+\nu$ events
than in generic hadronic events. 

The resulting branching fraction is

\begin{equation}
{\cal{B}}(D^+\to\mu^+\nu) = (4.40\pm0.66^{+0.09}_{-0.12})\times 10^{-4}.
\end{equation}

\noindent Using Eq.~\ref{eq:width} we determine the decay constant to be:

\begin{equation}
f_{D^+}=(222.6\pm16.7^{+2.8}_{-3.4})~{\rm MeV}.
\end{equation}

\section{Measurement of $f_{D_s^+}$}

	The measurements of $\fds$ at CLEO require higher energy running in order to produce
the $D_s\bar{D}_s$ pair. A scan of the energy region from 3970 to 4260 MeV was performed, and
it was determined that the optimal energy for $D_s$ physics was 4170 MeV~\cite{scan}, where 
$D_s\bar{D}_s^*$ is dominant, {\it e.g.,} $\sigma(D_s\bar{D}_s^*)=(916\pm50)$~pb and 
$\sigma(D_s\bar{D}_s^*)=(35\pm19)$~pb. A slight complication with using $D_s\bar{D}_s^*$ is
the additional ($\sim$150~MeV) photon(s) from the $D^*_s$ decay. Two independent analyses have
been carried out. The first analysis is similar to the $D^+\to\mu^+\nu$ measurement described previously,
where, in addition to measuring ${\cal{B}}(D_s^+\to\mu^+\nu)$, we also measure 
${\cal{B}}(D_s^+\to\tau^+\nu)$, where, $\tau^+\to\pi^+\nu\bar{\nu}$. In the second analysis,
we measure ${\cal{B}}(D_s^+\to\tau^+\nu)$, $\tau^+\to e^+\nu\bar{nu}$. 

\subsection{Measurement of ${\cal{B}}(D_s^+\to(\mu^+,\tau^+)\nu)$ using Missing Mass}

	We use 314~$\ipb$ of data collected at $E_{\rm cm}=4170$~MeV for this analysis.
We search for final states consistent with either $D_s^+\to\mu^+\nu$
or $D_s^+\to\tau^+\nu$. The branching fraction is obtained from:

\begin{equation}
{\cal{B}} = {N_{\rm cand} - N_{\rm back} \over N^*_{\rm tag}\epsilon}
\end{equation}

\noindent where $N^*_{\rm tag}$ is the number of reconstructed $D_sD^*_s$ events and
$\epsilon$ is the efficiency for reconstruction and identification of the $\mu^+$ for
$D_s^+\to\mu^+\nu$, or the $\pi^+$ for $D_s^+\to\tau^+\nu,~\tau^+\to\pi^+\nu\bar{nu}$
We therefore absorb the full reconstruction of the $D^*_s$ into the denominator, and
do not rely on Monte Carlo simulation for the efficiency of the $\sim$100~MeV photon.

\begin{figure}[h]
\centering
\includegraphics[width=80mm]{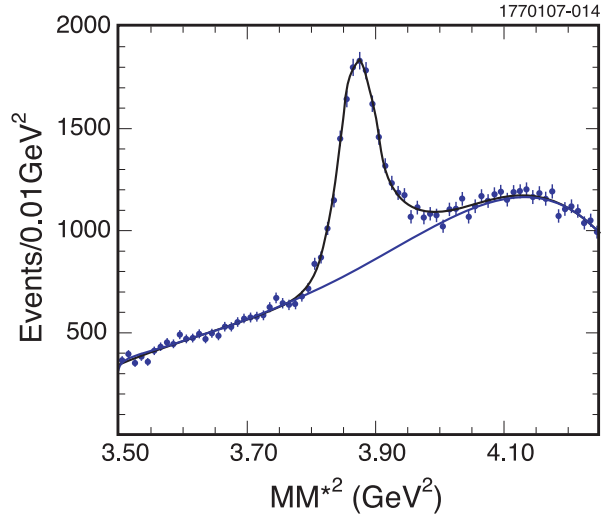}
\caption{Square of the missing mass recoiling against a $\gamma D^*_s$ candidates.}
\label{fig:mmst}
\end{figure}

	To determine $N^*_{\rm tag}$, we first fully reconstruct a hadronic $D_s^-$ tag in 
eight tag modes, from which we obtain $31,302\pm472$ $D_s^+$ tags. To identify $D_s\bar{D}^*+_s$ 
events, we combine a $D_s^-$ tag with any additional photon candidate in the event
and form the missing-mass squared ($MM^{*2}$) recoiling against the $\gamma D^+_s$ system, 
$MM^{*2}=(E_{\rm cm}-E_{D_s}-E_{\gamma})^2-(\vec{p}_{\rm cm}-\vec{p}_{D_s}-\vec{p}_{\gamma})^2$.
This quantity peaks at $M_{D_s}^2$, regardless of whether the photon came from the
$D_s^-$ (the tag) or from the $D_s^+$. The distribution of $MM^{*2}$ is shown in 
Fig~\ref{fig:mmst} for all eight tag modes combined. A fit to this distribution yields
$18645\pm426\pm1081$ $D^*_s\bar{D_s}$ events within $\pm$2.5 standard deviations of $M(D_s)$.

	To search for $D_s^+\to\mu^+\nu$ and $D_s^+\to\tau^+\nu,~\tau^+\to\pi^+\nu\bar{nu}$,
we require a single additional charged particle and no additional photon candidates with
energy in excess of 250 MeV. The signatures for $D_s^+\to\mu^+\nu$ and $D_s^+\to\tau^+\nu,~\tau^+\to\pi^+\nu\bar{nu}$
are similar in that they both have a $D_s$ tag and a single high momentum charged particle. In
addition to the difference in the energy depositions of muons and pions, the two-body versus
three-body decay implies significantly different missing mass ($MM^2$) distributions.  To 
suppress backgrounds with neutrals, we veto events which have an energy deposition (excluding 
the tag) in the $CC$ exceeding 250~MeV. The two-body
leptonic decay form a $MM^2$ distribution that peaks near zero with a resolution of $\sim$0.025~GeV$^2$. 
The three-body leptonic decay covers a broad $MM^2$ region, which peaks near 0.1 GeV$^2$, and 
falls smoothly to zero at $MM^2=-0.05$ and extends to $MM^2\sim0.8$~GeV$^2$. We thus define
signal samples as follows. (i)-$\mu$: For $D_s^+\to\mu^+\nu$, we require an energy deposition,
$\ecctrk<300$~MeV, and $|MM^2|<0.05$~GeV$^2$. For $D_s^+\to\tau^+\nu,~\tau^+\to\pi^+\nu\bar{\nu}$,
we define two subsamples -- (i)-$\tau$: $\ecctrk<300$~MeV and $0.05<MM^2<0.20$~GeV$^2$, and
(ii)-$\tau$ $\ecctrk>300$~MeV and $-0.05<MM^2<0.20$~GeV$^2$. The upper cutoff in $MM^2$ is to
avoid background from $D_s^+\to K^0\pi^+$. We also consider a third sample, (iii)-$e$, for
$D_s^+\to e^+\nu$ by requiring the track's energy deposition to be consistent with its momentum 
and $|MM^2|<0.050$~GeV$^2$. 

\begin{figure}[h]
\centering
\includegraphics[width=80mm]{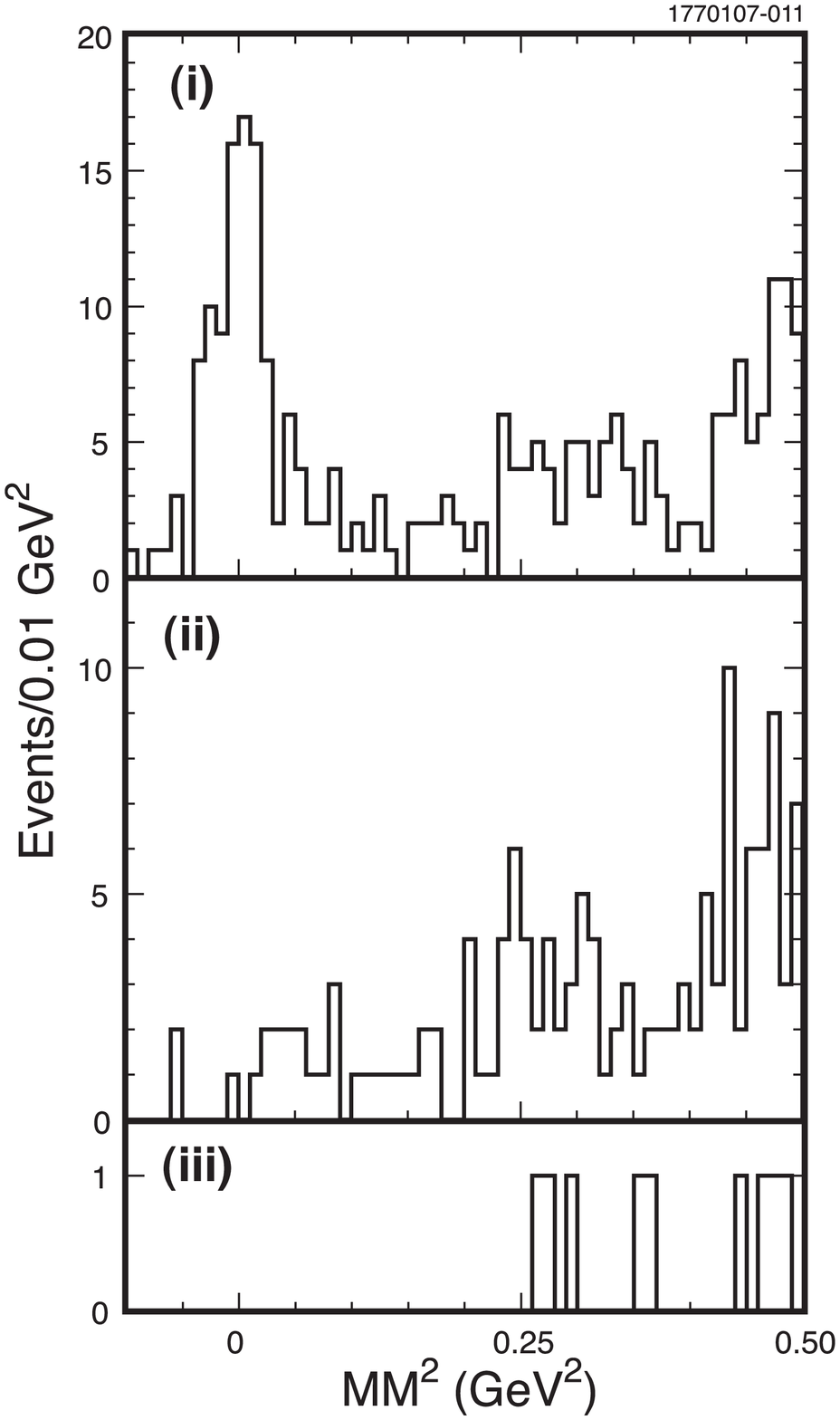}
\caption{Square of the missing mass recoiling against $\gamma D^*_s \mu^+ ({\rm or}~\pi^+)$ candidates
for cases (i) $\ecctrk<300$~MeV ($\dsmunu$ and $\dstaunu$ candidates combined), 
(ii) $\ecctrk>300$~MeV ($\dstaunu$), and (iii) charged particle consistent with an electron.}
\label{fig:mm}
\end{figure}

	The $MM^2$ distributions are shown in Fig.~\ref{fig:mm}, where 
cases (i)-$\mu$ and (i)-$\tau$ are combined. In the $\dsmunu$ signal region of
$|MM^2|<0.05$~GeV$^2$, we find 92 events
with an expected background of $3.5\pm1.4$ events. This sample is mostly of $\dsmunu$,
with some cross-feed from $\dstaunu$. We evaluate
the branching fraction using the number of $\mu^+\nu$ events, $N_{\mu\nu}$=92, in the signal 
region:

\begin{eqnarray}
N_{\mu\nu} &=& N_{\rm det}-N_{\rm back}  \\ \nonumber
           &=& N^*_{\rm tag}\cdot\epsilon[\ep\bf(\dsmunu) \\ \nonumber
& &    + \edp\bf(\dstaunu)],
\label{eq:nmunu}
\end{eqnarray}

\noindent where $\epsilon=80.1\%$ is the efficiency of reconstructing the charged particle
in a $\dsmunu$ event, and includes the veto on events with additional photons with $E>250$~MeV. The
quantity, $\ep=91.4\%$, is the product of the muon identification efficiency (99\%) and
the $MM^2<0.05$~GeV$^2$ requirement (92.3\%). The cross-feed efficiency, $\edp$=7.9\%,
which is the product of the efficiency of the pion depositing less than 300 MeV in the CC (60\%)
and the $MM^2<0.05$~GeV$^2$ requirement (13.2\%). One can re-express $\bf(\dstaunu$) as:

\begin{eqnarray}
\bf(\dstaunu) = R\cdot\bf(\tau^+\to\pi^+\nu) \nonumber \\
\times\bf(\dsmunu) = 1.059\cdot\bf(\dsmunu)
\end{eqnarray}

\noindent where we use the Standard Model ratio for R:

\begin{equation}
R = {\Gamma(D_s^+\to\tau^+\nu)\over \Gamma(\dsmunu)} = \left({\mt\over\mm}\right)^2 { \left(1-{\mt^2\over\mds^2}\right)^2 \over \left(1-{\mm^2\over\mds^2}\right)^2 }=9.72.
\label{eq:rval}
\end{equation}

\noindent We thus find:

\begin{equation}
\bf(\dsmunu) = (0.594\pm0.066\pm0.031)\%,
\end{equation}

\noindent where the 5.2\% systematic error is dominated by the 5\% uncertainty on $N^*_{\rm tag}$.

	We also compute $\bf(\dstaunu)$ using cases (i)-$\tau$ and (ii)-$\tau$. For
these two cases, we find yields of 31 and 25 events, and expected backgrounds of 
$3.5^{+1.7}_{-1.1}$ and 5.1$\pm$1.6 events, respectively. The fraction of $\dstaunu$ events in 
the respective $MM^2$ regions are 32\% and 45\%. We thus find:

\begin{equation}
\bf(\dstaunu) = (8.0\pm1.3\pm0.4)\%.
\end{equation}

With the measured branching fractions, $\bf(\dsmunu)$ and $\bf(\dstaunu)$, we measure
the ratio of partial widths, R=$13.4\pm2.6\pm0.2$ (defined in Eq.~\ref{eq:rval}), 
which is consistent with the Standard Model value of 9.72.

	We may improve on the precision of $\bf(\dsmunu)$ by combining the $\dsmunu$ and $\dstaunu$ 
candidates. We can still use Eq.~\ref{eq:nmunu}, except $\ep$ and $\edp$ increase from
91.4\%  and 7.9\% to 96.2\% and 45.2\%, respectively. We thus find an effective
branching fraction:

\begin{equation}
\bf^{\rm eff}(\dsmunu)=(0.638\pm0.059\pm0.033)\%.
\label{eq:bfeff}
\end{equation}

\noindent Again, the dominant systematic uncertainty (5\%) is on the number of $D_s^*$ tags.
	
\begin{figure}[h]
\centering
\includegraphics[width=80mm]{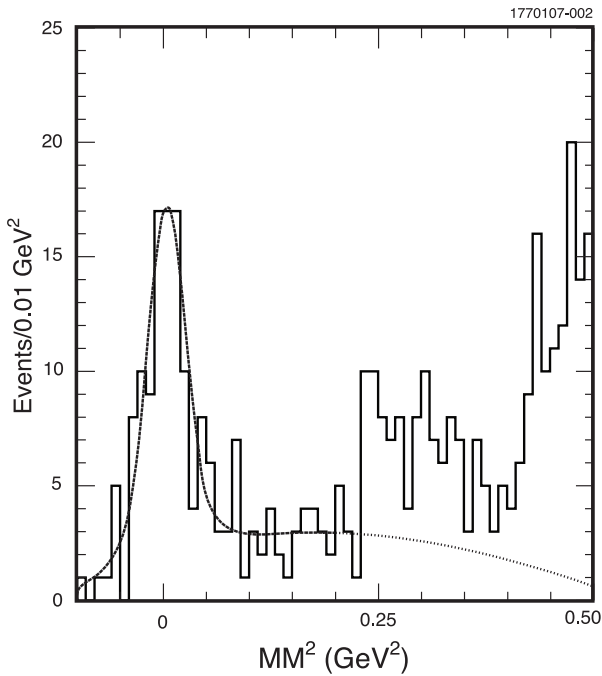}
\caption{Square of the missing mass recoiling against $\gamma D^*_s \mu^+ ({\rm or}~\pi^+)$ candidates.
The curve is the expected shape from simulation, normalized to the number of events with $MM^2<0.2$~GeV.}
\label{fig:mm2all}
\end{figure}

	The $MM^2$ distribution for all selected $\dsmunu$ and $\dstaunu$ candidates is shown
in Fig.~\ref{fig:mm2all}. Overlayed is a curve that represents the expected shape, normalized
to the event yield in the data in the $MM^2$ region below 0.2~GeV$^2$. We find good agreement between
the shape in data and expectations.

	We also search for the decay $D_s^+\to e^+\nu$. The helicity suppression in this
decay is much larger, and the expected rate is $\sim$50,000 times smaller than in $\dsmunu$. We find
no $D_s^+\to e^+\nu$ candidates and set the upper limit, $\bf(D_s^+\to e^+\nu)<1.3\times10^{-4}$ at 
the 90\% confidence level. 


	Using the more precise value for $\bf(\dsmunu)$ from Eq.~\ref{eq:bfeff}, we compute the
decay constant, $\fds$::

\begin{equation}
\fds = 274\pm13\pm7~{\rm MeV}
\label{eq:fds1}
\end{equation}

\noindent Combining this with our previous result for $\fd=(222.6\pm16.7^{+2.8}_{-3.4})$~MeV
we determine the ratio:

\begin{equation}
{\fds\over\fd} = 1.23\pm0.11\pm0.04.
\end{equation}

\section{Measurement of $\dstaunue$}

\begin{figure}[h]
\centering
\includegraphics[width=80mm]{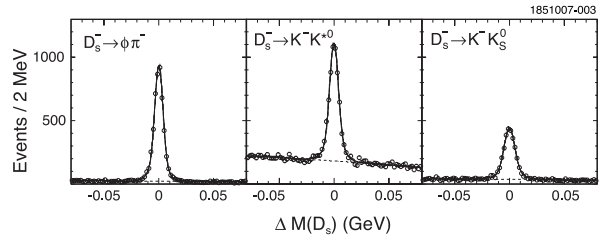}
\caption{Invariant mass distributions of $D_s^-$ candidates from data. The points are data,
the solid line is a fit, and the dashed line is the background.} 
\label{fig:dstag3}
\end{figure}

	In the second measurement of $\bf(D_s^+\to\tau^+\nu)$, we use 298~$\ipb$ of data
collected at $E_{\rm cm}=4170$~MeV. We utilize the decay $\tau^+\to e^+\nu\bar{\nu}$, where
we benefit from the large value of $\bf(\tau^+\to e^+\nu\bar{\nu})\sim18\%$, and the excellent
electron identification capabilities of the CLEO-c detector. We fully reconstruct the
three hadronic decay channels: $D_s^-\to\phi\pi^-, K^{*0}K^-$ and $K_S^0K^-$. 
Charged hadrons are identified using standard selection criteria~\cite{dhad}, and
the intermediate resonances, $\phi\to K^+K^-$, $K^{*0}\to K^-\pi^+$, and $K^0_S\to\pi^+\pi^-$,
are required to have an invariant mass within $\pm$10~MeV, $\pm$75~MeV and $\pm$12~MeV
of their known values~\cite{pdg}. Signal candidates are required to a reconstructed invariant mass, 
$M(D_s)$ within $\pm$20~MeV of the known $D_s$ mass ($m_{D_s}$). We also define sideband regions, 
$35<|M(D_s)-m_{D_s}|<55$~MeV, to study the combinatorial background. The invariant mass distributions
of the three $D_s^-$ tag channels are shown in Fig.~\ref{fig:dstag3}.

	To ensure we have $D_s\bar{D}_s^*$, we compute the mass recoiling against the reconstructed
$D_s$, and require it to be within $\pm$55~MeV of the $D_s^*$ mass~\cite{pdg}. We then select the subset of
events with a single additional charged track with $p>200$~MeV that has opposite charge to the 
$D_s$ tag and is consistent with being a positron. The discriminating variable we use to
identify $\dstaunue$ is $E_{\rm extra}$, the total energy remaining in the calorimeter after 
all showers associated with the tag and the positron are removed. In signal events, the only additional
particles beyond the $D_s$ tag and the positron are the two neutrinos and either
a photon from $D^*_s\to\gamma D_s$, or a $\pi^0$ from $D^*_s\to\pi^0 D_s$ .
Kinematically, these photons populate the energy regions from 114-170~MeV (for $\gamma D_s^+$)
and 39-117~MeV (from $\pi^0 D_s^+$).

\begin{figure}[h]
\centering
\includegraphics[width=80mm]{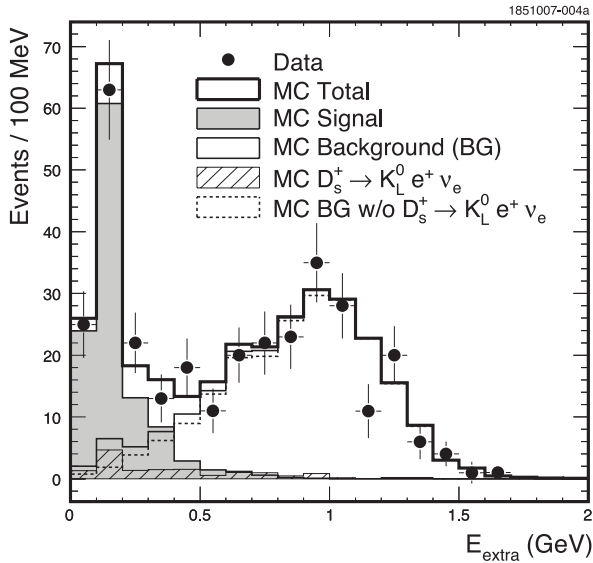}
\caption{Total extra energy left over in the calorimeter after removing energy associated with the
$D_s$ tag and the positron. Data are shown as the points with error bars, and the MC background
predictions are shown as solid, dashed and hatched histograms, and the expected signal 
contribution is indicated by the shaded histogram.}
\label{fig:eextra}
\end{figure}

	The distribution of $E_{\rm extra}$ in data is shown in Fig.~\ref{fig:eextra}.
The large excess at low values of $E_{\rm extra}$ is the $\dstaunue$ signal. The broad
background which peaks near 1~GeV is predominantly semi-leptonic decays, such as 
$D_s^+\to\phi e^+\nu,~\eta e^+\nu,~\etap e^+\nu,~K^0e^+\nu$ and $K^{*0}e^+\nu$. The
Cabibbo-suppressed decay, $K^0_Le^+\nu$, produces a small peaking component in the signal
region. The shape of this background is taken from Monte Carlo simulation, and is normalized 
to our measured rate for $D_s^+\to K^0_Se^+\nu$ of $\bf(D_s^+\to K^0_Se^+\nu)=(0.27\pm0.10)\%$.
We choose the signal region as $E_{\rm extra}<400$~MeV, which is chosen based on
optimizing the signal significance. The expected non-peaking
background in the signal region is estimated by scaling the number of data events with
$E_{\rm exta}>600$~MeV by the MC ratio of events in the sideband ($E_{\rm extra}^{\rm MC}>600$~MeV) to 
signal region ($E_{\rm extra}^{\rm MC}<400$~MeV). The yields of $D_s^-$ tags and 
$\dstaunue$ signal events are shown in Table~\ref{tab:sum2}. The scale factor, $s$ shown
in Table~\ref{tab:sum2} is a 
correction to account for slight differences in the expected number of background events
in the signal and sideband regions. Using the efficiency to reconstruct the final state,
$\dstaunue$ of $(71.4\pm0.4)\%$ and the $\bf(\tau^+\to e^+\nu\bar{\nu})=(17.84\pm0.05)\%$,
we find: 

\begin{equation}
\bf(D_s^+\to\tau\nu)=(6.24\pm0.71\pm0.36)\%. 
\end{equation}

\noindent The 5.8\% systematic uncertainty is dominated by the 4.3\% contribution from the 
simulation of $K_L^0$ showering in the calorimeter.

\begin{table*}[h]
\begin{center}
\caption{Summary of $D_s^-$ tagged events (yield, background from sidebands, sidebands scale factor ($s$),
and sideband-subtracted yield), and $\dstaunue$ events (yield, background from $D_s^-$ sidebands, 
background from $D_s^+$ semileptonic decays, and sideband-subtracted yield).}
\begin{tabular}{|l|c|c|c|c|c|c|c|c|}
\hline
~~~~~Mode               & \multicolumn{4}{|c|}{$D_s^-$ Tags} & \multicolumn{4}{|c|}{$\dstaunue$} \\ 
\cline{2-9} 
	 
                     & ~Yield~ &~Back~& $~s~$ &     Signal       & Yield &  \multicolumn{2}{|c|}{Background} & Signal \\
\cline{7-8} 
                     &        &      &   &                &       & ~~$D^-_s$~~ &   $D_s^+$     &          \\
\hline
$D_s^-\to\phi\pi^-$ &  5232  & 388  & 1.001 & $4843.6\pm75.0$  &  49   &   0     &   $8.7\pm0.6$ & $40.3\pm7.0$ \\
$D_s^-\to K^-K^{*0}$   &  8937  &3618  & 1.008 & $5289.2\pm112.2$ &  55   &   3     &   $8.5\pm0.7$ & $43.5\pm7.6$ \\
$D_s^-\to K^-K^0_S$ &  3468  & 695  & 1.030 & $2751.8\pm64.7$  &  24   &   2     &   $3.8\pm0.4$ & $18.1\pm5.1$ \\
\hline
Total                      &  17637 & 4701 &   -   & $12884.6\pm149.7$&  128  &   5     &   $21.0\pm1.0$& $101.9\pm11.5$ \\
\hline
\end{tabular}
\label{tab:sum2}
\end{center}
\end{table*}

	Using Eq.~\ref{eq:width}, we find $\fds=(275\pm16\pm8)$~MeV. When this result is combined with
the result in Eq.~\ref{eq:fds1}, we obtain:

\begin{equation}
\fds=274\pm10\pm5~MeV
\end{equation}

\noindent 

\section{Summary}

We have presented measurements of the branching fractions $\bf(D^+\to\mu^+\nu$, $\dsmunu$
and $\dstaunu$ with the CLEO-c detector. The results are the most precise measurements of these 
leptonic decay rates to date. Using Eq~\ref{eq:width}, we extract the decay constants:

\begin{eqnarray}
\fd &=& (222.6\pm16.7^{+2.8}_{-3.4})~{\rm MeV}.\\
\fds &=& (274\pm10\pm5)~{\rm MeV} \\
\fds/\fd &=& 1.23\pm0.11\pm0.03
\end{eqnarray}

	Our measurement of $\fds$ is consistent with and significantly more precise than
the recent measurement by BaBar~\cite{babar_fds}. The only other measurement of $\fd$
was reported by BES based on 1 signal candidate. Recent lattice QCD predictions~\cite{lqcd4,lqcd5} 
of both $\fd$ and $\fds$ are typically $\sim$10\% lower than our measurements, whereas the
ratio of $\fds/\fd$ is in good agreement with our measurement.

	We gratefully acknowledge the effort of the CESR staff in providing us with 
excellent luminosity and running conditions. We also thank the National Science Foundation
for support of this research.\newline

\end{document}